\documentclass[]{interact}

\usepackage{epstopdf}
\usepackage[caption=false]{subfig}

\usepackage[numbers,sort&compress]{natbib}
\bibpunct[, ]{[}{]}{,}{n}{,}{,}
\makeatletter
\def\NAT@def@citea{\def\@citea{\NAT@separator}}
\makeatother

\usepackage{graphics}
\usepackage{graphicx}
\usepackage{bm}
\usepackage{amsmath}
\usepackage{mathtools}
\usepackage{color}
\usepackage{bbold}
\usepackage{braket}

\begin{document}
\title{Locally-acting mirror Hamiltonians}
\author{\name{Jake Southall, Daniel Hodgson, Robert Purdy and Almut Beige}
\affil{The School of Physics and Astronomy, University of Leeds, Leeds LS2 9JT, United Kingdom}}

\maketitle

\begin{abstract}
Photons, i.e.~the basic energy quanta of monochromatic waves, are highly non-localised and occupy all available space in one dimension. This non-local property can complicate the modelling of the quantised electromagnetic field in the presence of optical elements that are local objects. Therefore, in this paper, we take an alternative approach and quantise the electromagnetic field in position space. Taking into account the negative- {\em and} the positive-frequency solutions of Maxwell's equations, we construct annihilation operators for highly-localised field excitations with bosonic commutator relations. These provide natural building blocks of wave packets of light and enable us to construct locally-acting interaction Hamiltonians for two-sided semi-transparent mirrors.
\end{abstract}

\section{Introduction}

In classical electrodynamics, we often characterise light by its local properties such as local amplitudes, direction of propagation and polarisation. Its fundamental equations of motion\textemdash Maxwell's equations\textemdash are local differential equations. Practically, we assume that the classical electromagnetic (EM) field comprises a continuum of local field excitations. In contrast to this, quantum electrodynamics routinely decomposes the EM field into monochromatic waves, which are highly non-local. Such a non-local approach can result in more complicated equations of motion than strictly necessary. For example, the Green's functions of macroscopic quantum electrodynamics correlate an observer's position with all spatial positions and photon frequencies \cite{Buhmann,Philbin,Buhmannx}. Therefore, in this paper, we take an alternative approach and quantise the EM field in position space. As in classical electrodynamics, our equations of motion only depend on local properties. Hence we expect them to find many applications, for example, in modelling systems involving local light-matter interactions or featuring ultrabroadband photonic wave packets \cite{exp,exp2,exp3,exp4}. 

To provide an example of a situation that we can analyse more easily in position than in momentum space, we focus in this paper on light scattering by two-sided semi-transparent mirrors. This topic already attracted a lot of interest in the literature (cf.~e.g.~Refs.~\cite{carniglia,Agarwal,in,in2,axel,Vogel,Hinds,Glauber,Barnett,Dalton,creatore,Nick,Ben}). In addition to using classical Green's functions \cite{Buhmann,Philbin,Buhmannx}, we can describe semi-transparent mirrors by restricting the Hilbert space of the EM field onto a subset of so-called triplet modes \cite{carniglia}. These consist of incident, reflected and transmitted waves and can be used to reproduce the well-known classical dynamics of field expectation values for light approaching a semi-transparent mirror from one side. However, they cannot describe situations in which wave packets approach a mirror surface from both sides without resulting in the prediction of unphysical interference effects \cite{Zako}. Some authors therefore prefer phenomenological approaches such as the input-output formalism \cite{in,in2,axel} or a quantum mirror image detector method that maps light scattering by semi-transparent mirrors onto analogous free-space scenarios \cite{Nick,Ben}. Although these models describe well the experiments that they have been designed for, they have not been derived from basic principles.

The mirror image method of classical electrodynamics simply describes light scattering by replacing any wave packet which comes in contact with the scattering object, at least partially, by its mirror image \cite{Jackson}. For semi-transparent mirrors, the mirror image is a wave packet with reduced field amplitudes which travels in the opposite direction and seems to emerge from the other side. In this paper, we take a similar approach. First, we quantise the EM field in position space and show that our approach includes the standard description of the EM field. Afterwards, using locally-acting field annihilation operators with bosonic commutator relations with respect to the conventional inner product, we construct locally-acting mirror Hamiltonians and show that these reproduce well-known classical dynamics. For example, they can cause a complete conversion of incoming into outgoing wave packets without altering the dynamics of outgoing wave packets. Since most quantum systems have a Hamiltonian, the same should apply to optical elements.

 When solving Maxwell's equations in free space in one dimension, we usually assume that their basic solutions are monochromatic travelling waves with real wave numbers $k$. By convention, positive and negative $k$ correspond to right- and left-moving wave packets, respectively. These monochromatic waves provide a complete description of the classical EM field, since they can be superposed to generate wave packets of any shape. For example, 
\begin{eqnarray} \label{walter5}
E(x) &=& {1 \over 2\pi} \int_0^\infty {\rm d} k \left[ E_0 \, {\rm e}^{{\rm i} k x}  + {\rm c.c.} \right]
\end{eqnarray}
are the electric field amplitudes of a highly-localised right-moving wave packet at position $x=0$ with amplitude $E_0$. To show that this is indeed the case, we substitute $k$ by $-k$ in the second term of the above equation and find that
\begin{eqnarray}  \label{walter6}
E(x,0) &=& {1 \over 2\pi} \int_{-\infty}^\infty {\rm d} k \, E_0 \, {\rm e}^{{\rm i} k x} ~=~  E_0 \, \delta(x) 
\end{eqnarray}
which is non-zero only at $x=0$. Next let us add an overall phase factor of ${\rm e}^{{\rm i}\pi/2} = {\rm i}$ to this highly-localised wave packet by adding a $\pi/2$ phase to all travelling waves. Doing so, the electric field amplitudes of the above wave packet become
\begin{eqnarray} \label{walter7}
E(x) &=& {1 \over 2\pi} \int_0^\infty {\rm d} k \left[ {\rm i} E_0 \, {\rm e}^{{\rm i} k x}  + {\rm c.c.} \right]
\end{eqnarray}
which no longer describes a highly-localised wave packet at $x=0$. Instead, the (real) electric field amplitudes are now given by 
\begin{eqnarray} \label{walter8}
E(x) &=& {{\rm i} E_0  \over 2 \sqrt{2} \pi} \left[ \int_0^\infty {\rm d} k \, {\rm e}^{{\rm i} kx } - \int_{-\infty}^0 {\rm d} k \, {\rm e}^{{\rm i} kx} \right] 
\end{eqnarray}
which is non-zero everywhere. This creates a problem, if we want to quantise the EM field in position space by associating the first term in Eq.~(\ref{walter5}) with the expectation value of an annihilation operator $a(x)$. As the above equations show, adding a factor ${\rm i}$ to $a(x)$ could change electric field amplitudes from being local to being non-zero everywhere. This should not be the case. 

As a solution, we include in the following both the positive- and the negative-frequency solutions of Maxwell's equations in our description of the EM field. More concretely, we assume that the basic solutions of Maxwell's equations are monochromatic travelling waves with positive and negative parameters $k$, two different directions of motion, $s = \pm 1$, and two different polarisations, $\lambda = {\sf H}, {\sf V}$. As above, these monochromatic travelling waves can be superposed to form wave packets of any shape. For example, 
\begin{eqnarray} \label{walter9}
E(x) &=& {1 \over 4 \pi} \int_{-\infty}^\infty {\rm d} k \left[ E_0 \, {\rm e}^{{\rm i} k x}  + {\rm c.c.} \right] \nonumber \\
&=& {1 \over 2} \left[ E_0  + {\rm c.c.} \right] \, \delta(x) \nonumber \\ 
&=& {\rm Re}(E_0) \, \delta(x)
\end{eqnarray}
describes a highly-localised wave packet at position $x=0$ with its electric field amplitude given by the real part of $E_0$. However, when associating the first term in this equation with the expectation value of an annihilation operator $a(x)$, it remains local when changing the relative phase of this operator, as is normally the case in quantum field theory.

The above described phase problem also seems to lie at the heart of the Fermi problem \cite{Fermi}. Indeed it has been shown that coupling two atoms to the same free radiation field with only positive-frequency photons results in a violation of Einstein causality \cite{Hegerfeldt,Milonni}, i.e.~the prediction of energy travelling from one atom to the other faster than allowed by the speed of light. Our intuition suggests that the dynamics of the atom-field system, which is caused by resonant and by off-resonant atom-field interactions, add complex phase factors to the coupling constants of the Hamiltonian in the interaction picture, thereby rendering it effectively non-local.

\begin{figure}[t]
\centering
\resizebox*{10cm}{!}{\includegraphics{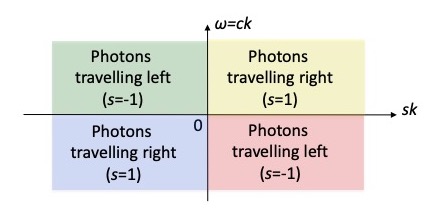}}
\caption{In this paper, we effectively double the usual Hilbert space of the quantised one-dimensional EM field and identify its basic energy quanta by their direction of motion $s = \pm 1$, their polarisation $\lambda = {\sf H}, {\sf V}$ and their frequency $\omega = ck$ which can be both positive and negative. In other words, as we shall see below, the eigenvalues $\hbar \omega$ of the Hamiltonian, which generates the dynamics of light, can be negative as well as positive. Using this notation, the always positive energy and corresponding classical wave number of a photon equal $\hbar c|k|$ and $sk$, respectively.} \label{map}
\end{figure}

Taking this into account, in this paper, we quantise both the negative- and the positive-frequency solutions of Maxwell's equations and assume that the dynamical Hamiltonian of the EM field has negative and positive eigenvalues $\hbar \omega$ with the photon frequency $\omega$ given by 
\begin{eqnarray} \label{walter10}
\omega &=& ck \, ,
\end{eqnarray}
where $c$ denotes the speed of light. As illustrated in Fig.~\ref{map}, this approach effectively doubles the Hilbert space of the quantised EM field compared to its standard description \cite{EJP}. Because of the above definition, we refer to photons with negative and positive $k$ in the following as negative- and positive-frequency photons, respectively. Using the notation in Fig.~\ref{map}, the wave number of a monochromatic travelling wave now equals $sk$. However, as we shall see below, the energy observable of the quantised EM field is positive and all photons have positive energy expectation values $\hbar c |k|$. Hence the dynamical Hamiltonian and the energy observable of the quantised EM field are no longer the same.

As we shall see below, considering both positive- and negative-frequency photons allows us to construct a local description of the quantised EM field which assumes that its basic building blocks are highly-localised field excitations. These are characterised by their respective position $x$, their direction of motion $s$ and their polarisation $\lambda$. Like all wave packets, highly-localised field excitations with a well-defined direction of motion travel at the speed of light which immediately implies an equation of motion. Later we see that this equation of motion can be cast into a Schr\"odinger equation. In this paper we aim for a position representation of the quantised EM field with a corresponding momentum representation and well-defined transformations between both \cite{Gray,Daniel}. Of course, these transformations must be reversible, which applies here due to the inclusion of the negative-frequency photons.

Many authors have attempted to introduce meaningful definitions of single-photon wave functions that are similar to the wave functions of massive particles \cite{Birula,Sipe,Birula2}. However, photon wave functions are a controversial concept since there are many differences between massive and massless particles. For example, this approach usually runs into problems because of the relativistic character of the free radiation field \cite{Cook,Cook2,Inagaki,Pavsic}. Another problem comes from the impossibility of fully localising a single-photon wave function in free space \cite{Knight,Eberly,Keller,Birula3}. As a result, many believe that it is impossible to quantise the EM field in position space and several no-go theorems have been put forward \cite{Wigner,Wightman,He,Malament}. Fortunately, these theorems are based on assumptions that can be overcome \cite{HB,HB2,HB3}. Like Hawton and Debierre \cite{Hawton,Hawton2}, we do this here by quantising the positive {\em and} the negative-frequency solutions of Maxwell's equations. In addition, we give up on the idea that the Hamiltonian and the energy observable of a system must always be the same.

However, the annihilation operators of highly-localised field excitations can be defined in different ways \cite{Daniel}. A concrete example of such annihilation operators is the $b$ operators of Titulaer and Glauber \cite{Glauber2} which describe the quantised EM field in terms of so-called temporal modes \cite{Raymer,Raymer2}. These do not obey bosonic commutator relations with respect to the conventional inner product and do not generate pairwise orthogonal states. Moreover, as we shall see below, it is possible to construct locally-acting annihilation operators {\em with} bosonic commutator relations. These can be used to construct locally-acting mirror Hamiltonians. By doing so, we identify further reasons for introducing negative-frequency photons which we summarise below:
\begin{enumerate}
\item Without the presence of the negative-frequency photons, we could not define annihilation operators for truly-localised field excitations with bosonic commutator relations which then allow us to construct locally-acting Hermitian mirror Hamiltonians.
\item Scattering by a semi-transparent mirror does not change the frequency of incoming photons. As we shall see below, positive frequencies remain positive and negative frequencies remain negative. 
As a result, it is possible to associate unitary beamsplitter transformations with effective interaction Hamiltonians, which only couple positive to positive-frequency photons. However, when neglecting the negative-frequency photon subspace, such an effective Hamiltonian always affects both incoming {\em and} outgoing wave packets. No distinction can be made between a wave packet travelling towards the mirror interface and one travelling away from it. 
\item Finally, as mentioned already above, we aim for a position representation of the quantised EM field with a corresponding momentum representation and well-defined transformations between the basis states of both descriptions. For more details, see Ref.~\cite{Daniel}.
\end{enumerate}

There are five sections in this paper. In Section \ref{Sec4}, we provide a complete description of the quantised EM field in terms of the annihilation and creation operators of highly-localised field excitations. This is done without specifying their commutator relations and without identifying their dynamical Hamiltonian. Section \ref{Sec2} studies the relation between the annihilation operators of highly-localised field excitations and the annihilation operators of the monochromatic excitations of the EM field. Moreover, we construct truly-local bosonic annihilation and creation operators with respect to the conventional inner product of quantum physics. In Section \ref{Sec3} we use these local bosonic operators to obtain locally-acting mirror Hamiltonians for modelling light scattering in the presence of two-sided semi-transparent mirrors and show, for example, that these Hamiltonians only affect incoming but not outgoing wave packets. Finally, we summarise our findings in Section \ref{Sec5}. 

\section{The position space representation of the quantised EM field} \label{Sec4}

In order to quantise the EM field in position space, we assume that the basic building blocks of wave packets of light are highly-localised field excitations. For simplicity, we only consider light propagating in one direction, i.e.~along the $x$-axis. Using the Heisenberg picture, we denote the annihilation operator of a highly-localised field excitation at position $x$ and at a time $t$ by $a_{s \lambda}(x,t)$. Here $\lambda = {\sf H},{\sf V}$ and $s=\pm 1$ refer to horizontally and to vertically polarised light, and to excitations propagating in the positive and the negative $x$-direction, respectively. As in classical electrodynamics, we demand that the expectation values of wave packets travel with the speed of light, $c$. In this paper, this is taken into account by assuming that 
\begin{eqnarray}
\langle \psi_{\rm H} | a_{s \lambda} (x,t) | \psi_{\rm H} \rangle  &=& \langle \psi_{\rm H} | a_{s \lambda} (x - sct,0) | \psi_{\rm H} \rangle 
\end{eqnarray}
for any state $| \psi_{\rm H} \rangle$ of the quantised EM field in the Heisenberg picture. Hence 
\begin{eqnarray} \label{oma}
a_{s \lambda}(x,t) &=& a_{s \lambda} (x - sct,0) \, .
\end{eqnarray}
This equation provides a fundamental equation of motion of the quantised EM field in free space. 

Next we notice that highly-localised field excitations are the origins of local field expectation values. In the following, we use Eq.~(\ref{oma}) to derive local field observables which are consistent with Maxwell's equations. In a medium with permittivity $\varepsilon$ and permeability $\mu$, where $c=1/\sqrt{\varepsilon \mu}$, and in the absence of any charges and currents, Maxwell's equations for the electric field $\mathbf{E} (\mathbf{r},t)$ and the magnetic field $\mathbf{B} (\mathbf{r},t)$ at positions $\mathbf{r}$ and at times $t$ are given by \cite{Nick}
\begin{eqnarray} \label{eq:maxwell}
&& \hspace*{-0.5cm} \nabla \cdot \mathbf{E} (\mathbf{r},t) = 0 \, , ~~ 
\nabla \times \mathbf{E} (\mathbf{r},t) = - \dot{\mathbf{B}} (\mathbf{r},t) \, , \nonumber \\
&& \hspace*{-0.5cm} \nabla \cdot \mathbf{B} (\mathbf{r},t) = 0 \, , ~~ 
\nabla \times \mathbf{B} (\mathbf{r},t) = \varepsilon  \mu \, \dot{\mathbf{E}} (\mathbf{r},t) \, .
\end{eqnarray}
The expectation values $\langle {\bf E}(x,t) \rangle$ and $\langle {\bf B}(x,t) \rangle $ of the observables ${\bf E}(x,t)$ and ${\bf B}(x,t)$ of the electric and the magnetic field in the Heisenberg picture need to solve these equations at all times. As we shall see below, one can show that this is indeed the case, if we assume that
\begin{eqnarray} \label{eq:1DEField}
{\bf E}(x,t) &=& \sum_{s = \pm 1} \sqrt{\hbar c \over \varepsilon A} \, \left[ \xi_{s {\sf H}} (x,t) \, \hat {\bf y} + \xi_{s {\sf V}} (x,t) \, \hat {\bf z} \right] \, , \nonumber \\
{\bf B}(x,t) &=& \sum_{s = \pm 1} {s \over c} \sqrt{\hbar c \over \varepsilon A} \, \left[ - \xi_{s {\sf V}} (x,t)  \, \hat {\bf y} + \xi_{s {\sf H}} (x,t) \, \hat {\bf z} \right] ~~~
\end{eqnarray}
with the operator $\xi_{s \lambda} (x,t)$ defined such that
\begin{eqnarray}
\xi_{s \lambda} (x,t) &= & {1 \over \sqrt{2}} \, [ a_{s\lambda}(x,t) + a^\dagger_{s\lambda}(x,t)] \, . 
\end{eqnarray}
Here $\hat {\bf y}$ and $\hat {\bf z}$ are unit vectors along the positive $y$ and $z$ axes and $A$ denotes the area around the $x$-axis which the EM field occupies. The normalisation factors in the above equation have been chosen here for convenience. Notice also that $\langle {\bf E}(x,t) \rangle$ and $\langle {\bf B}(x,t) \rangle $ are always real, since the $\xi_{s \lambda} (x,t)$ are Hermitian.

The chain rule of differentiation can be used to show that the time and the position derivative of $a_{s \lambda}(x-sct,0)$ are very closely linked,
\begin{eqnarray} \label{eq:A1}
{{\rm d} \over {\rm d}t} \, a_{s \lambda}(x-sct,0) &=& - sc \, {{\rm d} \over {\rm d}x}  a_{s \lambda}(x-sct,0) \, .
\end{eqnarray}
Combining this rule with Eqs.~(\ref{oma}) and (\ref{eq:1DEField}) yields 
\begin{eqnarray} \label{eq:A2}
\dot {\bf E}(x,t) &=& - \sum_{s = \pm 1} \, sc \, \sqrt{\hbar c \over 2 \varepsilon A} \, {{\rm d} \over {\rm d}x} \left[ a_{s {\sf H}} (x-sct,0) \, \hat {\bf y} +  a_{s {\sf V}} (x-sct,0)  \, \hat {\bf z} \right] + {\rm H.c.} \, , \nonumber \\
\dot {\bf B}(x,t) &=& - \sum_{s = \pm 1} \, \sqrt{\hbar c \over 2 \varepsilon A} \, {{\rm d} \over {\rm d}x} \left[ - a_{s {\sf V}} (x-sct,0) \, \hat {\bf y} + a_{s {\sf H}} (x-sct,0) \, \hat {\bf z} \right] + {\rm H.c.} 
\end{eqnarray}
Moreover, we see from Eq.~(\ref{eq:1DEField}) that
\begin{eqnarray} \label{eq:A3}
\nabla \times {\bf E}(x,t) &=& \sum_{s = \pm 1} \sqrt{\hbar c \over 2\varepsilon A} \, {{\rm d} \over {\rm d}x}\left[ - a_{s {\sf V}} (x-sct,0) \, \hat {\bf y} + a_{s {\sf H}} (x-sct,0) \, \hat {\bf z} \right] + {\rm H.c.} \, , \nonumber \\
\nabla \times {\bf B}(x,t) &=& - \sum_{s = \pm 1} {s \over c} \sqrt{\hbar c \over 2\varepsilon A} \, {{\rm d} \over {\rm d}x} \left[ a_{s {\sf H}} (x-sct,0) \, \hat {\bf y} + a_{s {\sf V}} (x-sct,0) \, \hat {\bf z} \right] + {\rm H.c.} ~~~~~
\end{eqnarray}
Comparing these equations with Eq.~(\ref{eq:maxwell}) confirms that $\langle {\bf E}(x,t) \rangle$ and $\langle {\bf B}(x,t) \rangle $ evolve indeed as predicted by Maxwell's equations.

From classical electrodynamics, we know that the observable $H_{\rm eng}(t)$ for the energy of the quantised EM field at time $t$ in free space must equal
\begin{eqnarray} \label{eq:H2}
H_{\rm eng}(t) &=& {A \over 2} \int_{-\infty}^{\infty} {\rm d} x \, \left[ \epsilon \, {\bf E}(x,t)^2 + {1 \over \mu} {\bf B}(x,t)^2 \right] .
\end{eqnarray}
Substituting ${\bf E}(x,t)$ and ${\bf B}(x,t)$ in Eq.~(\ref{eq:1DEField}) into this equation, one can show that 
\begin{eqnarray} \label{eq:H2app}
H_{\rm eng}(t) &=&  \sum_{s,s' = \pm 1} \sum_{\lambda = {\sf H}, {\sf V}} \int_{-\infty}^{\infty} {\rm d} x \,  {\hbar c \over 2} \, (1 + ss') \, \xi_{s\lambda}(x,t)  \xi_{s'\lambda}(x,t) \, .
\end{eqnarray}
Since only terms with $s' =s$ contribute to the above expression, it simplifies to
\begin{eqnarray} \label{final}
H_{\rm eng}(t) &=& \sum_{s = \pm 1} \sum_{\lambda = {\sf H}, {\sf V}} \int_{-\infty}^{\infty} {\rm d} x \, \hbar c \, \xi^\dagger_{s \lambda} ( x,t) \xi_{s \lambda} (x,t) \, . 
\end{eqnarray}
This equation shows that the expectation values of the energy of the quantised EM field are always positive. It also shows that $\xi^\dagger_{s \lambda} (x,t) \xi_{s \lambda} (x,t) $ is the observable for the energy density at position $x$ and time $t$. Moreover, Eq.~(\ref{final}) respects the translational symmetry of the quantised EM field. All inner degrees of freedom contribute equally to $H_{\rm eng}(t)$. It is worth noting here that, since $\xi_{s \lambda} (x,t)$ is Hermitian, it is not necessary to use the notation $\xi^\dagger_{s \lambda} (x,t)$. However, the dagger symbol helps to clarify the meaning of expressions.

The above description of the quantised EM field is complete in a sense that it provides expressions for all basic field observables, i.e.~the electric and magnetic field vectors and the field energy. In addition, Eq.~(\ref{oma}) provides us with an equation of motion which can be used to evolve state vectors and expectation values in time. In position space, the dynamics of wave packets is almost trivial since light simply travels at a constant speed. As mentioned already in the Introduction, we do not associate highly-localised field excitations with individual particles \cite{Birula}. Hence the commutator relations of the $a_{s \lambda}(x,t)$ operators are not known. As we shall see below, there are different ways of introducing locally-acting annihilation operators $a_{s \lambda}(x,t)$ without contradicting any of the above equations and while still taking the basic symmetries of the quantised EM field in free space into account. 

\section{Comparing different position and momentum space representations} \label{Sec2}

To learn more about the properties of the locally-acting annihilation operators $a_{s \lambda}(x,t)$, we now have a closer look at the corresponding momentum space representation of the quantised EM field in the Heisenberg picture. To do so, we introduce particle annihilation operators $a_{s \lambda}(k,t)$ with the bosonic commutator relations
\begin{eqnarray} \label{last200}
\big[ a_{s \lambda}(k,t), a^\dagger_{s' \lambda'}(k',t) \big] &=& \delta_{s,s'} \, \delta_{\lambda,\lambda'} \, \delta (k - k') \, . 
\end{eqnarray}
Inspired by classical electrodynamics and for reasons which will become more obvious below, we assume in the following that the $a_{s \lambda}(x,t)$ and the $a_{s \lambda}(k,t)$ operators are linked via a generalised Fourier transform,
\begin{eqnarray} \label{eq:U2}
a_{s \lambda}(x,t) &=& \int_{-\infty}^\infty {\rm d} k \, f(k) \, {\rm e}^{{\rm i}skx} \, a_{s \lambda} (k,t) \, .
\end{eqnarray}
Notice that the inverse transformation of this equation, which implies that
\begin{eqnarray} \label{eq:U22}
a_{s \lambda}(k,t) &=& {1 \over 2 \pi} \int_{-\infty}^\infty {\rm d} x \, f^{-1}(k) \, {\rm e}^{-{\rm i}skx} \, a_{s \lambda} (x,t) \, , ~~
\end{eqnarray}
only exists for any function $f(k)$ if both positive and negative-frequency photons are taken into account. This means that the parameter $k$ can assume any value between $- \infty$ and $+ \infty$.

From quantum field theory, we know that the quantised EM field is invariant under $\mathcal{PT}$ transformations, since light carries no internal quantum numbers \cite{Weinberg}. For example, in free space, a recording of the electric field amplitudes of a linearly-polarised wave packet which travels in the positive $x$-direction cannot be distinguished from a recording of the electric field amplitudes of an equally-shaped wave packet with the same polarisation which travels in the negative $x$-direction, if the second recording is played backwards in time \cite{schwinger,Joan,Joan2,vitiello}. This suggests that the equation for the electric field observable ${\bf E}(x,t)$ must be invariant under transformations which simultaneously replace $s$ and $t$ by $-s$ and $-t$. A similar argument can be made regarding the $\mathcal{PT}$ symmetry of the magnetic field observable ${\bf B}(x,t)$ in Eq.~(\ref{eq:1DEField}). 

However, an even stronger conclusion can be drawn from the equation of motion in Eq.~(\ref{oma}) while taking into account that 
\begin{eqnarray} \label{walter}
x - sct &=& x - (-s) c (-t) \, .
\end{eqnarray}
These two equations imply that the generators for the dynamics of the $a_{s \lambda}(x,t)$ operators and for the dynamics of the $a_{-s \lambda}(x,-t)$ operators are formally the same. Both must have the same set of eigenvalues. Hence for every wave packet which moves in the $s$ direction, there must be another wave packet which moves in the $-s$ direction which evolves in exactly the same way but with time running backwards. This implies that the dynamical Hamiltonian of the quantised EM field must have equal amounts of positive and negative eigenvalues. It also implies that the dynamical Hamiltonian $H_{\rm dyn}$ and the energy observable $H_{\rm eng}(t)$ in Eq.~(\ref{final}) of the quantised EM field, which only has positive eigenvalues, cannot be the same.

When combining the identity in Eq.~(\ref{oma}) with Eq.~(\ref{eq:U2}), we find that
\begin{eqnarray} \label{eq:U2t}
a_{s \lambda}(x,t) &=& \int_{-\infty}^\infty {\rm d} k \, f(k) \, {\rm e}^{{\rm i}sk(x-sct)} \, a_{s \lambda} (k,0) \, .
\end{eqnarray}
From quantum optics we know that these dynamics correspond to that of the harmonic oscillator Hamiltonian 
\begin{eqnarray} \label{eq:H22}
H_{\rm dyn} &=& \sum_{s = \pm 1} \sum_{\lambda = {\sf H}, {\sf V}} \int_{-\infty}^\infty {\rm d} k \, \hbar ck \, a^\dagger_{s \lambda}(k,0)  a_{s \lambda}(k,0) \, . 
\end{eqnarray}
This means that the states $|\psi_{\rm S}(t) \rangle$ of the quantised EM field in the Schr\"odinger picture evolve indeed according to a Schr\"odinger equation, 
\begin{eqnarray} \label{Terry}
{\rm i} \hbar \, |\dot \psi_{\rm S}(t) \rangle &=& H_{\rm dyn} \, |\psi_{\rm S}(t) \rangle \, .
\end{eqnarray}
The above dynamical Hamiltonian $H_{\rm dyn}$ applies for any choice of $f(k)$ and is almost the same as the usual harmonic oscillator Hamiltonian of the EM field in free space \cite{EJP}. However, $H_{\rm dyn}$ now has positive as well as negative eigenvalues and no longer coincides with $H_{\rm eng}(t)$, as mentioned already above. 

Next we have a closer look at the commutator relations of the annihilation operators $a_{s \lambda}(x,t)$ of highly-localised field excitations. For example, combining Eqs.~(\ref{last200}) and (\ref{eq:U2t}), one can show that
\begin{eqnarray} \label{last100}
 \big[ a_{s \lambda}(x,t), a^\dagger_{s' \lambda'}(x',t') \big] &=& \delta_{s,s'} \, \delta_{\lambda,\lambda'} \int_{-\infty}^\infty {\rm d} k \, |f(k)|^2 \, {\rm e}^{{\rm i} sk[ x- sct - (x'-sct')]}  \, . 
\end{eqnarray}
This commutator relation implies that 
\begin{eqnarray} \label{last}
\big[ \xi_{s \lambda}(x,t), \xi^\dagger_{s' \lambda'}(x',t') \big] &=& {\rm i} \,  \delta_{s,s'} \, \delta_{\lambda,\lambda'} \int_{-\infty}^\infty {\rm d} k \, |f(k)|^2 \, \sin \left[ sk( x- sct - (x' - sct')) \right] \, . \nonumber \\
\end{eqnarray}
As it should, this expression vanishes when $x-sct = x'-sct'$, $s=s'$ and $\lambda = \lambda'$, since $\xi_{s \lambda}(x,t)$ and $\xi_{s' \lambda'}(x',t')$ are identical in this case (cf.~Eq.~(\ref{oma})). 

Substituting Eq.~(\ref{eq:U2t}) into Eq.~(\ref{final}), we find that the energy observable $H_{\rm eng}(t)$ equals
\begin{eqnarray} \label{eq:H2app3}
H_{\rm eng}(t) &=& \sum_{s = \pm 1} \sum_{\lambda = {\sf H}, {\sf V}} \int_{-\infty}^{\infty} {\rm d} x \int_{- \infty}^{\infty} {\rm d} k \int_{- \infty}^{\infty} {\rm d} k' \,  {1 \over 2} \hbar c \nonumber \\
&& \times \Big[ f(k)^* f(k') \, {\rm e}^{-{\rm i}s(k- k')(x - sct)} \, a^\dagger_{s\lambda}(k,0) a_{s\lambda}(k',0) \nonumber \\ 
&& + f(k) f(k') \, {\rm e}^{{\rm i}s(k+ k')(x-sct)} \, a_{s\lambda}(k,0) a_{s\lambda}(k',0) + {\rm H.c.} \Big] 
\end{eqnarray}
in momentum space. Performing the $x$ integration yields $\delta$-functions. Taking this into account, the energy observable of the EM field eventually simplifies to 
\begin{eqnarray} \label{eq:H3}
H_{\rm eng}(t) &=& \sum_{s = \pm 1} \sum_{\lambda = {\sf H}, {\sf V}} \int_{-\infty}^\infty {\rm d} k \, \pi \hbar c \, \Big[ \, |f(k)|^2 \, a^\dagger_{s\lambda}(k,0) a_{s\lambda}(k,0) + {\rm H.c.} \nonumber \\
&& + f(k) f(-k) \, a_{s\lambda}(k,0) a_{s\lambda}(-k,0) + {\rm H.c.} \Big] ~~~~
\end{eqnarray} 
which is independent of time, as it should be. The expectation values of the terms in the second line of this equation vanish if photon states with only positive or only negative $k$'s are populated. 

Next we examine the extra terms in Eq.~(\ref{eq:H3}) which make the energy observable different from an harmonic oscillator Hamiltonian. As usual, the vacuum state $|0 \rangle$ is the shared zero eigenstate of all photon annihilation operators $a_{s \lambda}(k,t)$. It describes an EM field with zero energy and zero electric and magnetic field expectation values. As we can see from Eq.~(\ref{eq:U2}), the vacuum state is also annihilated by all locally-acting annihilation operators $a_{s \lambda}(x,t)$ and 
\begin{eqnarray} \label{vac}
a_{s \lambda}(k,t) \, |0 \rangle = a_{s \lambda}(x,t) \, |0 \rangle &=& 0 
\end{eqnarray}
for all parameters $k$, $x$ and $t$. Moreover, the vacuum state $|0 \rangle$ is an example of the coherent states $|\alpha_{s \lambda} (k) \rangle$ with
\begin{eqnarray}
a_{s \lambda} (k,0) \, |\alpha_{s \lambda} (k) \rangle &=& \alpha_{s \lambda}(k) \, |\alpha_{s \lambda} (k) \rangle 
\end{eqnarray}
which we parametrise as usual by complex numbers $\alpha_{s \lambda} (k)$. Considering Eqs.~(\ref{eq:1DEField}) and (\ref{eq:U2}), one can show that there are different coherent states  with the same field expectation values $\langle {\bf E} (x,t) \rangle $ and $\langle {\bf B} (x,t) \rangle$. For example, for real $f(k)$, this applies to the coherent states $|\alpha_{s \lambda} (k) \rangle$ and $|\alpha_{s \lambda} (-k) \rangle$ with $\alpha_{s \lambda} (-k) = \alpha_{s \lambda} (k)^*$. Both coherent states describe light travelling in the same direction and it is impossible to distinguish them by looking only at their electric and magnetic field expectation values. Hence the electric and magnetic field amplitudes of a state of the form $|\alpha_{s \lambda} (k) \rangle |\alpha_{s \lambda} (-k) \rangle $ with $\alpha_{s \lambda} (-k) = \alpha_{s \lambda}(k)^*$ interfere constructively. Their total energy $\langle H_{\rm eng}(t) \rangle$ is therefore four times as large as the energy of $|\alpha_{s \lambda} (k) \rangle$ (cf.~Eq.~(\ref{eq:H3})). Analogously, one can show that states of the form $|\alpha_{s \lambda} (k) \rangle |\alpha_{s \lambda} (-k) \rangle $ with $\alpha_{s \lambda} (-k) = -\alpha_{s \lambda}(k)^*$ have the same energy expectation value $\langle H_{\rm eng}(t) \rangle = 0$ as the vacuum state. This is why the energy observable no longer coincides with a harmonic oscillator Hamiltonian. 

\subsection{Highly-localised field excitations} \label{subsec}

As mentioned already above, there are many different consistent choices for the function $f(k)$ which correspond to different physical descriptions of the quantised EM field \cite{Daniel}. One possibility is to assume that  
\begin{eqnarray} \label{eq:U2f}
f(k) &=& \sqrt{|k| \over 2 \pi} \, {\rm e}^{{\rm i} \, {\rm sgn}(k)\phi} \, ,
\end{eqnarray}
where $\phi $ with $\phi \in [0,2\pi)$ is a free parameter. In this case, the above model of the quantised EM field becomes co-variant and the fields transform as expected under Lorentz transformations \cite{Daniel}. The factor sgn$(k)$ denotes the sign of $k$ and has been added to the above exponent to ensure that $H_{\rm eng}(t)$ in Eq.~(\ref{eq:H3}) remains independent of $\phi$. Calculating this energy observable for the above choice of $f(k)$ shows that the $a_{s \lambda}(k)$ can be interpreted as the annihilation operators of monochromatic photons of energy $\hbar c |k|$, and the locally-acting annihilation operators $a_{s \lambda}(x,0)$ have many similarities with the $b$ annihilation operators of Titulaer and Glauber \cite{Glauber2}. 

However, the above $|f(k)|$ is non-zero not only for positive but also for negative $k$ values. This means, in this paper we have effectively doubled the Hilbert space of the quantised EM field compared to its standard description which only considers positive $k$'s. This change of Hilbert space also affects the commutator relations of the EM field. Now the last term in Eq.~(\ref{last}) vanishes and
\begin{eqnarray} \label{last2}
\big[ \xi_{s \lambda}(x,t), \xi^\dagger_{s' \lambda'}(x',t') \big] &=& 0 \, .
\end{eqnarray}
This equation implies that electric and magnetic fields can be measured independently everywhere. For travelling waves, local electric and magnetic field amplitudes are essentially the same\textemdash they only differ by a constant factor (cf.~Eq.~(\ref{eq:1DEField})). It is therefore not surprising that the observables ${\bf E}(x,t)$ and ${\bf B}(x,t)$ commute. However, authors who only quantise a subset of all available standing waves after imposing certain boundary conditions usually obtain a different commutator \cite{Buhmann,Philbin,Buhmannx,commutator} which illustrates the incompleteness of their description.

The excited states of the quantised EM field in position space in the Schr\"odinger picture are obtained by applying creation operators $a^\dagger_{s \lambda} (x,0)$ to the vacuum state. For example, the (unnormalised) state
\begin{eqnarray}
|1_{s \lambda} (x) \rangle &=& a^\dagger_{s \lambda} (x,0) \, |0 \rangle
\end{eqnarray}
describes a single highly-localised field excitation at position $x$. When calculating the overlap between two single-excitation states, we see that this overlap depends on the commutator relation of the respective local annihilation and creation operators. More concretely, we find that
\begin{eqnarray}
\langle 1_{s \lambda} (x) |1_{s' \lambda'} (x') \rangle &=& \langle 0| \big[ a_{s \lambda} (x,0), a^\dagger_{s' \lambda'} (x',0) \big] |0 \rangle 
\end{eqnarray}
without any approximations. Substituting Eqs.~(\ref{last100}) and (\ref{eq:U2f}) into this equation eventually yields
\begin{eqnarray} \label{last500}
\langle 1_{s \lambda} (x) |1_{s' \lambda'} (x') \rangle &=& {1 \over 2 \pi} \, \delta_{s,s'} \, \delta_{\lambda,\lambda'} \int_{-\infty}^\infty {\rm d} k \, |k| \, {\rm e}^{{\rm i} sk (x- x')} \, .
\end{eqnarray}
This equation can be linked to the spatial derivative of a $\delta$-function \cite{Leonhardt}. For example, one can show that 
\begin{eqnarray} \label{last500}
\langle 1_{s \lambda} (x) |1_{s \lambda} (0) \rangle &=& {1 \over 2 \pi} \int_{-\infty}^\infty {\rm d} k \, |k| \, {\rm e}^{{\rm i} sk x} = {s \over \pi} \, {{\rm d} \over {\rm d} x} \, {\rm Im} \left( \int_0^\infty {\rm d} k \, {\rm e}^{{\rm i} sk x} \right) \, . ~~~~~
\end{eqnarray}
Strictly speaking, the above expression is non-zero for all positions $x$ which means that there is a non-zero probability to detect the state $|1_{s,\lambda}(0) \rangle$ anywhere along the $x$-axis. However, the above equation also shows that this probability is extremely small unless $x$ is very close to $x=0$. This means, for most purposes it is well justified to associate the $a_{s \lambda}(x,t)$ operators with $f(k)$ as in Eq.~(\ref{eq:U2f}) with highly-localised field excitations.

\subsection{Truly-local bosonic field excitations}

However, as we shall see below, for certain applications like the construction of locally-acing interaction Hamiltonians, it is useful to introduce annihilation operators for truly-localised field excitations. To do so, we now define annihilation operators $A_{s \lambda}(x,t)$ with $f(k)$ in Eq.~(\ref{eq:U2}) such that
\begin{eqnarray} \label{eq:U2ff}
f(k) &=& {1 \over \sqrt{2 \pi}} \, {\rm e}^{{\rm i} \, {\rm sgn}(k)\phi} \, .
\end{eqnarray}
Using this equation, one can show that the annihilation operators $A_{s \lambda}(x,0)$ obey bosonic commutator relations,
\begin{eqnarray} \label{last600}
\big[ A_{s \lambda}(x,0), A^\dagger_{s' \lambda'}(x',0) \big] &=& \delta_{s,s'} \, \delta_{\lambda,\lambda'} \, \delta (x-x') \, . ~~
\end{eqnarray}
Hence the overlap of different single-excitation states $|1_{s \lambda} (x) \rangle$ with
\begin{eqnarray}
|1_{s \lambda} (x) \rangle &=& A^\dagger_{s \lambda} (x,0) \, |0 \rangle
\end{eqnarray}
is simply given by 
\begin{eqnarray} \label{last700}
\langle 1_{s \lambda} (x) |1_{s' \lambda'} (x') \rangle &=& \delta_{s,s'} \, \delta_{\lambda,\lambda'} \, \delta (x-x') \, . ~~
\end{eqnarray}
The single-excitation states of the $A_{s \lambda}(x,0)$ operators are pairwise orthogonal and the field excitations created when applying $A^\dagger_{s \lambda}(x,0)$ operators to the vacuum state are therefore truly-localised.  

As we have seen above, in general, there is no need to distinguish between positive and negative-frequency photons when modelling light propagation in free space \cite{Cook,Cook2,Inagaki}. However, to obtain locally-acting annihilation operators with bosonic commutator relations, the coefficients $f(k)$ must be non-zero for positive and for negative $k$ values. The price we pay for this extension of the standard Hilbert space of the quantised EM field is that the representation of field observables, like the electric and magnetic field vectors ${\bf E}(x,t)$ and ${\bf B}(x,t)$, is more complicated when using the $A_{s \lambda}(x,t)$ operators than when using the $a_{s \lambda}(x,t)$ operators but it is not impossible. For more details see Ref.~\cite{Daniel}.

\subsection{How to return to the standard description of the quantised EM field}

If we want to recover the usual textbook expressions for the electric and magnetic field observables, ${\bf E}(x) = {\bf E}(x,0)$ and ${\bf B}(x) = {\bf B}(x,0)$ of the quantised EM field \cite{EJP}, we need to choose $\phi = \pi / 2$ and $f(k) = 0$ for $k < 0$, while assuming that $f(k)$ is as given in Eq.~(\ref{eq:U2f}) for $k>0$. In this case, the energy observable $H_{\rm eng} (t)$ in Eq.~(\ref{eq:H3}) simplifies to the dynamical Hamiltonian $H_{\rm dyn}$ in Eq.~(\ref{eq:H22}) and the $a_{s \lambda}(k,t)$ operators all describe photons with positive frequencies $\omega = ck$ and positive energies $\hbar \omega$. As we have seen in Section \ref{subsec}, while the $a_{s \lambda} (k,t)$ obey bosonic commutator relations, the commutator relations of the corresponding annihilation operators $a_{s \lambda} (x,t)$ with $f(k)$ as described above are non-trivial.

\section{Two-sided semi-transparent mirrors} \label{Sec3}

In classical electrodynamics, we usually use a mirror image method and local electric and magnetic field vectors to model light scattering by mirror interfaces \cite{Jackson}. Suppose a wave packet travels along the $x$ axis towards a mirror which has been placed in the $x=0$ plane. In general, incoming and outgoing wave packets simply evolve as they would in free space. However, once an incoming wave packet comes in contact with the mirror surface, it is replaced by its mirror image. The mirror image is a wave packet with negative electric field amplitudes that travels in the opposite direction and seems to originate from the opposite side of the mirror. In the case of a semi-transparent mirror, the conversion of incoming into outgoing wave packets is incomplete and the amplitude of the reflected light is reduced by the respective reflection rate.

In this section, we model the dynamics of the quantised EM field near a two-sided semi-transparent mirror in an analogous fashion. As in classical electrodynamics, we describe any incoming light in position representations. To reproduce the above-described impact of the mirror interface, we construct locally-acting mirror Hamiltonians. As we shall see below, these replace any incoming wave packet by its mirror image but do not affect wave packets far away from the interface. Depending on the size of the interaction constants of the mirror Hamiltonian, the resulting scattering transformation can be complete or incomplete. 

For many applications, for example, for the modelling of beam splitters in linear optics experiments, it is enough to know the overall scattering transformations of incoming wave packets. These can be deduced from the well-known classical dynamics of incoming wave packets while imposing unitary transformation operators and assuming energy conservation. In Section \ref{scatter}, we see that our approach allows us to derive such scattering transformations from basic principles. Moreover, locally-acting mirror Hamiltonians can be used to study the dynamics of incoming wave packets during the scattering process in detail. This is illustrated in Section \ref{appC}, where we consider a concrete example. However, in contrast to other methods (e.g.~Ref.~\cite{carniglia}), the main advantage of our approach is that it can describe semi-transparent mirrors with light approaching not only from one but from both sides.

Let us begin by noticing that the presence of a mirror should not affect our notion of the basic energy quanta of the quantised EM field, since it does not change the nature of incoming and outgoing wave packets. It only changes how these wave packets evolve in time. In the following we therefore include the presence of a mirror interface in our description of light propagation by altering the relevant system Hamiltonian. Taking this approach already worked very well when describing light scattering through two-sided optical cavities \cite{JMO}. More concretely, we are looking for a Hermitian Hamiltonian of the form 
\begin{eqnarray} \label{HHH}
H_{\rm mirr} &=& H_{\rm dyn} + H_{\rm int}
\end{eqnarray}
which 
\begin{enumerate}
\item is time-independent in the Schr\"odinger picture;
\item preserves the energy of any incoming light; 
\item acts locally and only affects wave packets in contact with the mirror interface, i.e.~which does not affect light that is moving away from the mirror surface; 
\item reproduces the well-known scattering dynamics of light in the presence of two-sided semi-transparent mirrors and can reverse the direction of incoming wave packets;
\item preserves the orbital angular momentum of the incoming light by transforming circular-polarised light into circular-polarised light of the same type.
\end{enumerate}
To construct such a mirror Hamiltonian we first return into the Schr\"odinger picture, where all field annihilation and creation operators are time-independent. To distinguish these operators from the Heisenberg operators $A_{s \lambda}(x,t)$ and $A^\dagger_{s \lambda}(x,t)$ which we considered in the previous two sections, we add a superscript ${\rm (S)}$ and define
\begin{eqnarray}  \label{standard}
A^{(\rm S)}_{s \lambda}(x) &=& A_{s \lambda}(x,0) \, .
\end{eqnarray}
Next we define bosonic annihilation operators $A^{(\rm S)}_{s \pm}(x)$ for truly localised field excitations of circular-polarised light, 
\begin{eqnarray}  \label{standard}
A^{(\rm S)}_{s \pm}(x) &=& {1 \over \sqrt{2}} \, \left[ A^{(\rm S)}_{s {\sf H}}(x) \pm {\rm i} \, A^{(\rm S)}_{s {\sf V}}(x) \right] \, .
\end{eqnarray}
Using this notation, a mirror interaction Hamiltonian $H_{\rm int}$ which obeys all of the above conditions and respects basic thermodynamical principles for the construction of interactions \cite{Stokes} is given by 
\begin{eqnarray}  \label{eq:H7}
H_{\rm int} &=& \sum_{\lambda = \pm} \int_{-\infty}^\infty {\rm d}x \int_{-\infty}^\infty {\rm d}x' \, {\rm i} \hbar \Omega_{xx'} \, \left[ A^{(\rm S)}_{1\lambda} (x) \, A^{(\rm S) \dagger}_{-1 \lambda} (x') - {\rm H.c.} \right] 
\end{eqnarray}
with the $\Omega_{xx'}$ denoting real coupling constants. These must be non-zero only near the mirror interface and zero everywhere else.

At any given time $t$, the above interaction annihilates truly-localised field excitations at positions $x$ and replaces them with truly-localised field excitations at $x'$ which travel in the opposite direction. Adjusting its coupling constants $\Omega_{xx'}$ accordingly, $H_{\rm int}$ can describe a mirror interface of any thickness and with different material properties. For simplicity, we consider in the following a relatively thin mirror near the $x=0$ plane. In this case, the coupling constants $\Omega_{xx'}$ are only non-zero for $x$ and $x'$ close to the origin of the $x$-axis. Since $\Omega_{xx'}$ is a number and not an operator, we describe the effect of the mirror interface with the help of a local classical potential \cite{Hofmann}. This should be well justified, if incoming wave packets collide with a macroscopic collection of coherent and freely-moving electrons inside the mirror interface such that their orbital angular momentum is conserved, while the direction of propagation is reversed. 

\subsection{The interaction picture} \label{IP}

Before analysing the dynamics associated with the mirror Hamiltonian $H_{\rm mirr}$ in Eq.~(\ref{HHH}), let us move from the Schr\"odinger into the interaction picture with respect to the free Hamiltonian $H_0 = H_{\rm dyn}$ and with respect to $t=0$. In the following, $|\psi_{\rm I}(t) \rangle$ denotes the state vector of the quantised EM field at time $t$ in the interaction picture with $|\psi_{\rm S}(t) \rangle$ being the corresponding state vector in the Schr\"odinger picture. As usual in physics, we define the state vector in the interaction picture such that
\begin{eqnarray}
|\psi_{\rm I}(t) \rangle &=& U_{\rm dyn}^\dagger (t,0) \, |\psi_{\rm S}(t) \rangle \, . 
\end{eqnarray}
Taking the time derivative of the above equation, one can show that this state vector evolves according to a Schr\"odinger equation but with the corresponding Hamiltonian given by 
\begin{eqnarray} \label{E1}
H_{\rm I}(t) &=& U_{\rm dyn}^\dagger (t,0) \, H_{\rm int} \, U_{\rm dyn} (t,0) \, . ~~~
\end{eqnarray}
In the absence of a mirror potential, $|\psi_{\rm I}(t) \rangle = |\psi_{\rm S}(0) \rangle$ at all times and local field excitations remain at their initial positions. The purpose of changing into the interaction picture is to simplify the following calculations by removing all free-space dynamics from the time evolution of the EM field. 

Since the dynamical Hamiltonian moves wave packets at the speed of light along the $x$ axis, we know that
\begin{eqnarray} \label{oma556}
U^\dagger_{\rm dyn}(t,0) \, A^{(\rm S)}_{s \lambda} (x) \, U_{\rm dyn} (t,0) &=& A^{(\rm S)}_{s \lambda}(x - sct) \, .
\end{eqnarray}
Hence moving the mirror Hamiltonian $H_{\rm mirr}$ in Eq.~(\ref{HHH}) into the interaction picture yields the interaction Hamiltonian
\begin{eqnarray} \label{E1}
H_{\rm I}(t) &=& \sum_{\lambda = \pm} \int_{-\infty}^\infty {\rm d}x \int_{-\infty}^\infty {\rm d}x' \, {\rm i} \hbar \Omega_{xx'} \left[ A^{(\rm S)}_{1 \lambda} (x-ct) A^{(\rm S) \dagger}_{-1 \lambda} (x'+ct) - {\rm H.c.} \right] . ~~~
\end{eqnarray}
Next we substitute $\tilde x = x - ct$ and $\tilde x' = x' + ct$, which simplifies the above equation to
\begin{eqnarray} \label{E10}
H_{\rm I}(t) &=& \sum_{\lambda = \pm} \int_{-\infty}^\infty {\rm d}\tilde x \int_{-\infty}^\infty {\rm d}\tilde x' \, {\rm i} \hbar \Omega_{(\tilde x+ct)(\tilde x'-ct)} \left[  A^{(\rm S)}_{1 \lambda} (\tilde x) A^{(\rm S) \dagger}_{-1 \lambda} (\tilde x') - {\rm H.c.} \right] . ~~~
\end{eqnarray}
In the interaction picture, the mirror potential travels at the speed of light away from its original position. Later on in Section \ref{appC}, where we have a closer look at a concrete example, we see that we can now analyse the dynamics of incoming wave packets at all times $t$ and at any position $x$, if the coupling constants $\Omega_{xx'}$ are fully known.

\subsection{Incoming versus outgoing wave packets} \label{IVA}

However, let us first have a closer look at the general properties of locally-acting mirror Hamiltonians. In this subsection, we show that the above introduced mirror potential is only seen by incoming wave packets, while remaining invisible to outgoing wave packets. To show that this is indeed the case, let us assume for a moment that the mirror is placed in the $x=0$ plane and that the coupling constants $\Omega_{xx'}$ are essentially non-zero only when both $x$ and $x'$ are very close to the origin of the $x$ axis. In this case, as we have seen in the previous subsection, incoming wave packets correspond in the interaction picture to local field excitations with $x<0$ and $s=1$ or with $x>0$ and $s=-1$. Moreover, outgoing wave packets correspond to local field excitations with $x<0$ and $s=-1$ or with $x>0$ and $s=-1$.

Suppose a truly-localised right-moving field excitation is placed on the left-hand side of the mirror interface at a position $\tilde x < 0$ at $t=0$. In this case, it reaches the interface and experiences the mirror potential after a time $t = |\tilde x|/c$. At this point, $\tilde x + ct$ equals zero and the coupling constants $\Omega_{(\tilde x+ct)(\tilde x'-ct)}$ in Eq.~(\ref{E10}) become non-zero for $\tilde x' = |\tilde x|$. Hence the interaction Hamiltonian $H_{\rm I}(t)$ annihilates the incoming excitation at $\tilde x$ and replaces it with a truly-localised left-moving field excitation at $\tilde x' = - \tilde x$. which is on the opposite side of the mirror interface. This is exactly what one would expect in the interaction picture according to the predictions of the mirror image method of classical electrodynamics \cite{Jackson}. In contrast to this, a left-moving wave packet on the left hand side of the mirror never reaches a position where $\Omega_{(\tilde x+ct)(\tilde x'-ct)}$ differs from zero. It therefore never experiences the mirror interaction and remains at its original location. The same applies to a right-moving wave packet on the right hand side of the mirror interface. 

\subsection{The scattering transformation of incoming wave packets} \label{scatter}

Since this too can be done without specifying the mirror interaction constants $ \Omega_{xx'} $, we now derive the overall scattering operator $S_{\rm I}$,
\begin{eqnarray} \label{D1app2}
S_{\rm I} &=& \exp \left( - {{\rm i} \over \hbar} \int_{-\infty}^\infty {\rm d}t \, H_{\rm I}(t) \right) ,
\end{eqnarray}
for light scattering by a semi-transparent mirror in the interaction picture. 
Combining Eqs.~(\ref{eq:U2}) and (\ref{eq:U2ff}) and having a closer look at the annihilation operators $A^{(\rm S)}_{s \lambda}(x)$ of truly-localised field excitations, we find that
\begin{eqnarray} \label{eq:U2zz}
A^{(\rm S)}_{s \lambda}(x) &=& {1 \over \sqrt{2 \pi}} \int_{-\infty}^\infty {\rm d} k \, {\rm e}^{{\rm i} \, {\rm sgn}(k)\phi} \, {\rm e}^{{\rm i} skx} \, a_{s \lambda} (k,0) \, . \end{eqnarray}
Choosing $\phi = 0$ for convenience and substituting this equation into Eq.~(\ref{E1}) yields the interaction Hamiltonian 
\begin{eqnarray} \label{D1app}
H_{\rm I}(t) &=& { {\rm i} \hbar \over 2\pi}  \sum_{\lambda = \pm} \int_{-\infty}^\infty {\rm d}x \int_{-\infty}^\infty {\rm d}x' \int_{- \infty}^\infty {\rm d} k \int_{- \infty}^\infty {\rm d} k' \, \Omega_{xx'} \nonumber \\
&& \times \Big[ {\rm e}^{{\rm i} (kx + k'x')} \, {\rm e}^{-{\rm i} (k - k')ct} \, a_{1 \lambda} (k,0) \, a^\dagger_{-1 \lambda} (k',0) - {\rm H.c.} \Big] \, . 
\end{eqnarray}
Using this Hamiltonian and performing the time integration in Eq.~(\ref{D1app2}), which can be done without knowing the coupling constant $\Omega_{xx'}$, yields a $\delta$-function in momentum space. When subsequently performing the $k'$ integration, only terms with $k' = k$ contribute and the above scattering operator $S_{\rm I}$ simplifies to
\begin{equation} \label{D1app3}
S_{\rm I} = \exp \left( - {\rm i} \sum_{\lambda = \pm} \int_{- \infty}^\infty {\rm d} k \left[ \Xi_k \, a_{1 \lambda} (k,0) \, a^\dagger_{-1 \lambda} (k,0) + {\rm H.c.} \right] \right)  
\end{equation}
with the $k$-dependent complex coupling constants $\Xi_k $ defined as
\begin{eqnarray} \label{oma5}
\Xi_k &=& {{\rm i} \over c} \int_{-\infty}^\infty {\rm d}x \int_{-\infty}^\infty {\rm d}x' \, \Omega_{xx'} \, {\rm e}^{{\rm i} k(x + x')} \, .
\end{eqnarray}
The above scattering operator has been derived without approximations. Moreover, we see that it couples positive- to positive- and negative- to negative-frequency photons. It also preserves energy and couples circular-plus to circular-plus and circular-minus to circular-minus polarised light, as it should. Moreover, for coupling constants $\Xi_k= \pi/2$, the above scattering operator $S_{\rm I}$ results in a complete transfer of excitation from a $(1,k)$ mode into a $(-1,k)$ mode and vice versa.

The observation that there is no mixing between positive-and negative-frequency photons suggests that it is possible to construct an effective mirror Hamiltonian for positive-frequency photons without considering negative-frequency photon operators. This is almost the case. Suppose an interaction Hamiltonian 
\begin{eqnarray} \label{D1app4}
H_{\rm I} (t) &=& \sum_{\lambda = \pm} \int_0^\infty {\rm d} k \, \left[ \hbar \Omega_k \, a_{1 \lambda} (k,0) a^\dagger_{-1 \lambda} (k,0) + {\rm H.c.} \right] 
\end{eqnarray} 
couples left- to right- and right- to left-moving photons with $\Omega_k$ denoting the respective coupling constants. When calculating the scattering operator $S_{\rm I}$ associated with this Hamiltonian, we find that it has the same effect within the $k>0$ subspace as $S_{\rm I}$ in Eq.~(\ref{D1app3}), if the coupling constants $\Omega_k$ can be chosen such that 
\begin{eqnarray} \label{D1app4xx}
\int_{- \infty}^\infty {\rm d} t \,  \Omega_k &=& \Xi_k    
\end{eqnarray}
with $\Xi_k$ defined as in Eq.~(\ref{oma5}). Unfortunately, for finite $\Omega_k$, this is not possible, since the integral on the left hand side of this equation is always infinitely large. Moreover, the effective Hamiltonian $H_{\rm I} (t)$ in Eq.~(\ref{D1app4}) always affects incoming as well as outgoing wave packets. This means, considering only the standard Hilbert space of the quantised EM field with $k>0$, we cannot specify the position of the mirror interface. Obtaining a complete description of the quantised EM field requires a doubling of its standard Hilbert space, as has previously been proposed in Refs.~\cite{Nick,Ben}. 

\subsection{A concrete example of a locally-acting mirror Hamiltonian} \label{appC}

Having a closer look at Eq.~(\ref{oma5}) suggests that the main contribution to the scattering dynamics of incoming wave packets comes from the $\Omega_{xx'}$ terms in the mirror interaction Hamiltonian with $x'=-x$. Contributions with $x' \neq -x$ average away when we perform the $k$ integration and contribute less significantly to the dynamics of incoming wave packets. Keeping this in mind, we assume in the following that 
\begin{eqnarray} \label{C2}
\Omega_{xx'} &=& \Omega(x) \, \delta(x+x') 
\end{eqnarray}
as a concrete example. In this case, the Hamiltonian $H_{\rm int}$ in Eq.~(\ref{eq:H7}) describes an interaction which transforms truly-localised field excitations at position $x$ with local (real) coupling strength $\Omega(x)$ into truly-localised field excitations at $-x$. For example, for very narrow mirror potentials with $\Omega_{xx'} \neq 0$ only when $x$ and $x'$ are both very close to the origin of the $x$ axis, the assumption that $x' \approx -x$ is in general well justified (both variables are essentially zero) and the above mirror potential always applies to a very good approximation. 
 
Substituting Eq.~(\ref{C2}) into Eq.~(\ref{E10}) and performing the $x'$ integration simplifies the mirror interaction Hamiltonian $H_{\rm I}(t)$ such that 
\begin{eqnarray} \label{E1x}
H_{\rm I}(t) &=& \sum_{\lambda = \pm} \int_{-\infty}^\infty {\rm d}\tilde x \, {\rm i} \hbar \Omega(\tilde x+ct) \, \left[  A^{(\rm S)}_{1 \lambda} (\tilde x) A^{(\rm S) \dagger}_{-1 \lambda} (-\tilde x) - {\rm H.c.} \right] . 
\end{eqnarray}
Ehrenfest's theorem tells us that the time derivative of $\langle A \rangle_t = \langle \psi_{\rm I}(t) | A |\psi_{\rm I}(t) \rangle$, i.e.~of the expectation value of a time-independent operator $A$ of a given state $|\psi_{\rm I}(t) \rangle$, equals
\begin{eqnarray} \label{E1xx}
\langle \dot A \rangle_t &=& - {{\rm i} \over \hbar} \langle [ A , H_{\rm I}(t) ] \rangle_t 
\end{eqnarray}
in the interaction picture. Using this equation and employing the bosonic commutator relations of the truly-local annihilation operators to calculate the time derivatives of the expectation values of $ A^{(\rm S)}_{1 \lambda} (x)$ and $ A^{(\rm S)}_{-1 \lambda} (-x)$, we find that
\begin{eqnarray} \label{D1app7}
\langle \dot A^{\rm (S)}_{1 \lambda} (x) \rangle_t &=& - \Omega (x+ct) \, \langle A^{(\rm S)}_{-1 \lambda}(-x) \rangle_t \, , \nonumber \\
\langle \dot A^{\rm (S)}_{-1 \lambda} (-x) \rangle_t &=& \Omega (x+ct) \, \langle A^{(\rm S)}_{1 \lambda}(x) \rangle_t \, .
\end{eqnarray}
Solving these linear differential equations for known expectation values at $t=0$ yields 
\begin{eqnarray} \label{D1app77}
\left( \begin{array}{c} \langle A^{\rm (S)}_{1 \lambda} (x) \rangle_t  \\ \langle A^{\rm (S)}_{-1 \lambda} (-x) \rangle_t \end{array} \right)
&=& \left( \begin{array}{rr}  \cos \left( \Xi(x,t) \right) & - \sin \left( \Xi(x,t) \right) \\ \sin \left( \Xi(x,t) \right) & \cos \left( \Xi(x,t) \right) \end{array} \right)
\left( \begin{array}{c} \langle A^{\rm (S)}_{1 \lambda} (x) \rangle_0  \\ \langle A^{\rm (S)}_{-1 \lambda} (-x) \rangle_0 \end{array} \right) ~~
\end{eqnarray}
with the real parameter $\Xi(x,t)$ given by
\begin{eqnarray} \label{D1app8}
\Xi(x,t) &=&  \int_0^t {\rm d} t' \, \Omega (x + ct') \, .
\end{eqnarray}
Returning into the Schr\"odinger picture with the help of the equations in Section \ref{IP} shows how light is scattered by a two-sided semitransparent mirror interface given the coupling constants in Eq.~(\ref{C2}). 

Suppose the coupling constants $\Omega(x)$ differ from zero only for positions $x$ very close to zero. In this case, for a given time interval $(0,t)$, the above equations only affect right-moving field excitations at positions $x \in (- ct,0)$ and left-moving field excitations at positions $-x \in (0,ct)$, as one would expect. Moreover, we see that, in the interaction picture, the time evolution annihilates truly-localised field excitations at positions $x$ and replaces them with truly-localised field excitations at positions $-x$, as the mirror potential swipes past. During reflection, the shape of an incoming wave packet remains exactly the same. It differs only by an overall phase factor, a reduction in its field amplitudes, and the reversal of its direction of propagation. The corresponding dynamics of incoming wave packets in the Schr\"odinger picture is therefore exactly what one would expect from the mirror-image method of classical electrodynamics \cite{Jackson}.

How much light is reflected and how much light is transmitted depends on the interaction strength of the mirror interface. For certain values of the coupling constant $\Xi(x,t) $ in Eq.~(\ref{D1app8}) with $t = \infty$, the above model describes a complete conversion of incoming into outgoing wave packets. This applies for example when $\Xi(x,\infty) = \pi/2 $. However, in general, mirror interfaces preserve the shape of incoming wave packets only to a very good approximation and corrections must be taken into account. For $\Omega_{xx'} \neq 0$ also when $x' \neq -x$, an incoming truly-localised field excitation spreads out and the scattering dynamics of incoming wave packets becomes more complex.

\section{Conclusions} \label{Sec5}

In this paper, we assume that the natural basic building blocks of light are highly-localised field excitations, since these can be combined easily into spread-out wave packets. We then introduce locally-acting annihilation operators $a_{s \lambda} (x,t)$ to describe light, which travels along the $x$ axis, in the Heisenberg picture. Here $x$, $s$ and $\lambda$ specify the location, direction of propagation and polarisation of highly-localised field excitations. Like all wave packets, highly-localised field excitations move with the speed of light. This observation implies a basic equation of motion (cf.~Eq.~(\ref{oma})). Together with the symmetries of the quantised EM field in position space, this equation provides an ideal starting point for a more detailed analysis. For example, it is shown that the quantum states of light evolve with a Schr\"odinger equation. However, the corresponding dynamical Hamiltonian $H_{\rm dyn}$ no longer coincides with the positive-definite field energy observable $H_{\rm eng}(t)$. Nevertheless, our description also contains states that evolve according to the standard description of the quantised EM field in momentum space \cite{EJP}. For more details see Ref.~\cite{Daniel}.

Highly-localised field excitations are the origin of local electric and magnetic field amplitudes. Therefore, we can use their annihilation operators to construct the electric and magnetic field observables ${\bf E}(x,t)$ and ${\bf B}(x,t)$ in the Heisenberg picture. In doing so, we obtain commuting expressions which shows that electric and magnetic fields can be measured simultaneously everywhere. This is not surprising since the local electric and magnetic field amplitudes of travelling waves only differ by a constant factor but are, otherwise, essentially the same. Moreover, we find that the annihilation operators $a_{s \lambda} (x,t)$ can be written as superpositions of annihilation operators $a_{s \lambda} (k,t)$ with bosonic commutator relations. However, the transformation between position and momentum space representations is only reversible if we quantise the negative- as well as the positive-frequency solutions of Maxwell's equations (cf.~Fig.~\ref{map}). 

In this paper, we motivated the introduction of negative-frequency photons by showing that this approach enables us to construct locally-acting mirror Hamiltonians $H_{\rm mirr}$ that reproduce the well-known dynamics of wave packets in the presence of two-sided semi-transparent mirrors. To do so, we define truly-local field annihilation operators $A_{s \lambda}(x,t)$ with bosonic commutator relations. These are a special example of the $a_{s \lambda}(x,t)$ operators of highly-localised field excitations and can be used to create pairwise orthogonal states in position space. In addition to constructing locally-acting Hamiltonians, these operators naturally lend themselves to the modelling of the quantised EM field in inhomogeneous media, are likely to provide new insight into fundamental quantum effects (cf.~e.g.~Refs.~\cite{Fermi,Hegerfeldt,Milonni,Unruh}) and into quantum information processing with photonic wave packets \cite{exp4}. 

When analysing the dynamics of the quantised EM field in the presence of a mirror interface in the interaction picture, we find that the scattering operator $S_{\rm I}$ in Eq.~(\ref{D1app3}), which maps the states of incoming onto the states of outgoing wave packets, does not change the frequency $\omega = ck$ of incoming photons. This shows that describing overall scattering transformations does not require an extension of the standard description of photonic wave packets. However, positive- and negative-frequency photons need to be taken into account if one wants to obtain a mirror Hamiltonian that only affects incoming but not outgoing wave packets. 

\section*{Acknowledgements}

JS and AB acknowledge financial support from the Oxford Quantum Technology Hub NQIT (grant number EP/M013243/1). We thank B. Granet, A. Kuhn, J. K. Pachos and N. Furtak-Wells for many stimulating and inspiring discussions. Statement of compliance with EPSRC policy framework on research data: This publication is theoretical work that does not require supporting research data.

\end{document}